\documentclass[12pt]{emulateapj}
\slugcomment{{\em}}
\usepackage{epsf}

\begin{document}

\title{Properties of Early-Type, Dry Galaxy Mergers and the Origin of Massive Elliptical Galaxies}

\author{Thorsten Naab$^1$, Sadegh Khochfar$^2$ \& Andreas Burkert$^1$}
\affil{$^1$ Universit\"ats-Sternwarte M\"unchen, Scheinerstr.\ 1, D-81679 M\"unchen, 
Germany; \texttt{naab@usm.lmu.de, burkert@usm.lmu.de} \\
$^2$ Department of Physics, Denys Wilkinson Bldg., University
of Oxford, Keble Road, Oxford, OX1 3RH, UK; \texttt{sadeghk@astro.ox.ac.uk}}

\begin{abstract}
The luminosity dependence of kinematical and isophotal properties of elliptical
galaxies is investigated using numerical simulations of galaxy merging,
combined with semi-analytical models of hierarchical structure formation.
Mergers of spiral galaxies as the only formation mechanism for elliptical 
galaxies can neither reproduce the kinematical and photometric properties of very
 massive elliptical galaxies nor the change from rotationally 
flattened disky to anisotropic boxy systems with increasing luminosity. We present 
numerical simulations showing that binary mergers of early-type galaxies open an 
additional channel for 
the formation of anisotropic, slowly rotating and boxy ellipticals. 
Including this channel in a semi-analytical model we can successfully reproduce 
the observed trend that more luminous giant ellipticals are more boxy and less flattened 
by rotation. This trend can be strengthened by suppressing residual gas infall and star 
formation for galaxies with stellar bulge masses $M_* \geq 3 \times 10^{10} M_{\odot}$. 
Hence we propose that mergers of early-type galaxies play an important role for the 
assembly of massive elliptical galaxies.  
\end{abstract}


\section{Introduction}
\begin{figure*}
\epsscale{1.}
\plotone{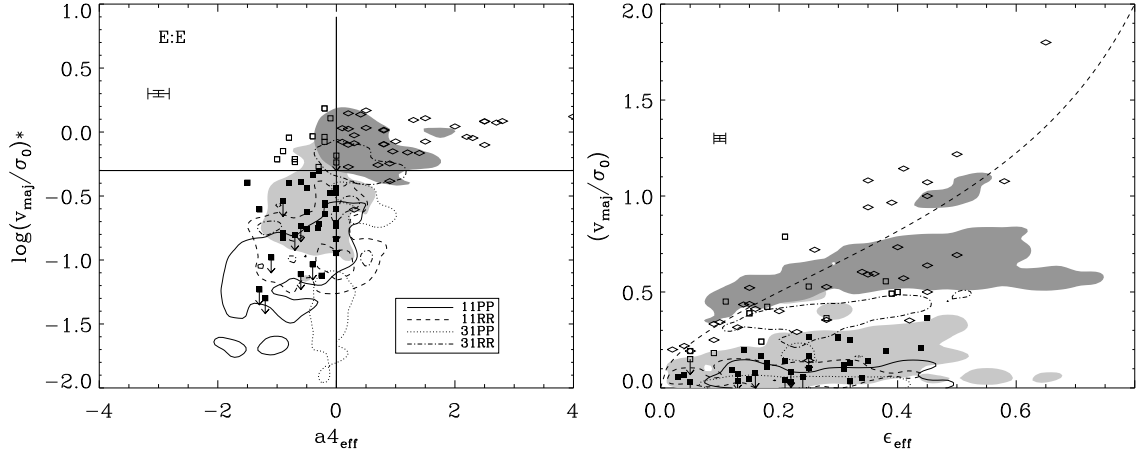}
\caption{{\it Left panel}: Isophotal shape $a4_{\mathrm{eff}}$ versus anisotropy 
parameter $(v_{\mathrm{maj}}/\sigma_0)^*$. The contours indicate the 80\% 
probability to find a remnant of 1:1 anisotropic (11AA; solid), 
1:1 rotationally flattened  
(11RR; dashed), 3:1 anisotropic (31AA; dotted), 
and 3:1 rotating (31RR; dashed-dotted) early-type mergers in the enclosed 
area. For every initial galaxy pair the remnants of the three 
adopted impact parameters have been combined. The horizontal line at 
$(v_{\mathrm{maj}}/\sigma_0)^* =0.5$ separates the anisotropic boxy 
from the more rotationally flattened disky and boxy ellipticals. 
{\it Right panel}: Ellipticity $\epsilon_{\mathrm{eff}}$ versus the ratio of 
major axis rotation velocity and central velocity dispersion $(v_{maj}/\sigma_0)$.   
For comparison, in both panels the location of 1:1 (bright grey) and 
3:1 (dark grey) spiral merger remnants are indicated by the shaded areas 
(NB03). The data for observed disky (diamonds), very anisotropic 
boxy (filled boxes) and less anisotropic boxy (open boxes) ellipticals have 
been kindly provided by Ralf Bender. 11AA, 11RR, and 31AA remnants 
are consistent with slowly rotating, anisotropic and boxy 
elliptical galaxies. \label{fig1}}
\end{figure*}

Giant elliptical galaxies show a strong dependence of their kinematical and 
photometric properties with luminosity (see e.g. \citealp{1996ApJ...464L.119K}). 
Traditionally the kinematics supporting the shape of ellipticals is parametrized 
by the anisotropy parameter $({v_{\mathrm{maj}}/\sigma_0})^*$
which relates the observed ratio of the velocity along the major axis and the central 
velocity dispersion, ${v_{\mathrm{maj}}/\sigma_0}$, to the value expected for an isotropic 
rotator ${v/\sigma} = \sqrt{\epsilon/(1-\epsilon)}$  with the observed ellipticity $\epsilon$
\citep{1978MNRAS.183..501B}.  Observed slow rotating $(v_{\mathrm{maj}}/\sigma_0 \le 0.2)$ 
ellipticals in general have $(v_{\mathrm{maj}}/\sigma_0)^* \le 0.5$ (see Fig.\ref{fig1}) 
and their shape is assumed to be supported by anisotropic velocity dispersions. In the following we 
call these systems anisotropic (A). They tend to be the brightest 
ellipticals with  boxy isophotes and have the oldest, most metal rich stellar 
populations. The bulk of their stars formed on very short 
time-scales (see e.g. \citet{2004Natur.428..625H,2005ApJ...621..673T} 
and references therein). Low luminosity ellipticals rotate faster 
$(v_{\mathrm{maj}}/\sigma_0 > 0.2)$ and have $(v_{\mathrm{maj}}/\sigma_0)^* > 0.5$. 
They are assumed to be flattened by additional rotation (R), have disky (sometimes boxy) 
isophotes, younger stellar populations, and formed on longer time-scales. 
Recently, observations as well as simulations have shown that disky ellipticals show a  
significant underlying velocity anisotropy 
\citep{2005MNRAS.363..937B,2005MNRAS.363..597B,2005astro.ph..9470C}. In this sense  
remnants with $(v_{\mathrm{maj}}/\sigma_0)^* > 0.5$ might be partly supported by 
anisotropic dispersions and partly by intrinsic rotation. 

What is the origin of the strong mass dependence of the properties of elliptical galaxies? 
Does it fit into a scenario where ellipticals result 
from major galaxy mergers? 
While the properties of low luminosity ellipticals are consistent 
with being remnants of early-type spiral mergers 
(\citealp{2003ApJ...597..893N}(NB03); \citealp{2005astro.ph..8362N}) it is for several reasons   
unlikely that high luminosity ellipticals have formed in a similar way:
(i) the stellar populations of spirals are too young and metal poor and their
formation timescales are too long 
(see e.g. \citealp{2005ApJ...625...23B,2004Natur.428..625H} 
and references therein), (ii) very luminous ellipticals are significantly 
more massive than typical spirals, (iii) numerical simulations
indicate that kinematic and photometric properties of spiral merger remnants 
disagree with the properties of the most massive ellipticals 
(NB03; \citealp{2005astro.ph..8362N}). 
Although equal-mass spiral-spiral mergers can form slowly rotating, boxy remnants 
\citep{1999ApJ...523L.133N} they do, in general, not result in a homogenous 
family similar to bright ellipticals.     
  
\citet{2003ApJ...597L.117K} and \citet{2005MNRAS.359.1379K} (hereafter KB05) used 
semi-analytical modeling to demonstrate that massive ellipticals should have assembled 
predominantly in mergers between bulge-dominated gas poor (dry) early-type galaxies. 
They concluded that the observed characteristic 
properties of massive ellipticals might reflect the physics of early-type mergers
which in this case should lead to predominantly
boxy and slowly rotating, anisotropic systems, independent of the progenitor
mass ratio. This would be in contrast to spiral-spiral mergers which produce
boxy, anisotropic systems only if the progenitors have equal masses. 
The semi-analytical models of KB05 predict that early-type mergers also 
occur at low redshifts. These merger remnants would still appear old if gas 
infall and subsequent star formation was negligible during 
and after the merger events.

Some numerical simulations agree with the conclusions
of KB05. Gas-free mergers of spheroids preserve the 
fundamental plane \citep{2003MNRAS.342..501N,2003MNRAS.342L..36G,2005astro.ph..2495B}
and the properties of the remnants are in general similar to observed ellipticals 
\citep{2000ASPC..197..267N,2005MNRAS.361.1043G}. However, a detailed 
investigation of the isophotal shape and kinematics of 
dry, early-type merger remnants was missing up to now.

There is direct observational evidence for the existence of 
early-type mergers in galaxy clusters 
\citep{1999ApJ...520L..95V,2005astro.ph..5355T}.  \citet{2005astro.ph..6425B} 
and \citet{2005astro.ph..6661V} also find clear evidence for this process in the field. 
Comparing with merger timescales taken from the simulations presented here 
\citet{2005astro.ph..6425B} conclude that a 
spheroidal galaxy should have undergone typically 0.5-2
early-type major mergers since redshift $z\approx 0.7$. 
Investigations of the evolution of the luminosity function of elliptical 
galaxies using various surveys show that the number of early-type galaxies 
with luminosities 0.5-2 $L_*$ has been growing since  
a redshift of $z\approx 1$ accompanied by an increase of the total stellar 
mass in the early-type galaxy population 
\citep{2004ApJ...608..752B,2005astro.ph..6044F}. These observations support the 
scenario that massive galaxies assemble 
a significant amount of their mass even at $z < 1$ by major mergers of early-type 
galaxies and accretion of satellites rather than by star formation 
\citep{2004ApJ...608..742D,ks05}. 

In this letter we present the kinematic and photometric properties of simulated 
early-type merger remnants as a function of their initial conditions. 
We combine the outcome of our simulations with semi-analytical models and 
predict the present day abundance of anisotropic and 
rotationally flattened  ellipticals. 

\section{The merger models}
\label{MODELS}
As prototypical progenitors for mergers between early type galaxies we use an 
anisotropic remnant of a spiral-spiral merger with a mass ratio 1:1 and a 
rotationally flattened remnant of a 3:1 spiral-spiral 
merger from the sample of NB03. These objects resemble low and intermediate mass 
elliptical galaxies with respect to their
photometric and kinematic properties 
(NB03; \citealp{2005astro.ph..8362N}) and therefore 
can be considered as very good, but not perfect models for early-type galaxies.  
However, the mismatch in higher order moments of the line-of-sight velocity 
distribution of 3:1 remnants and observed ellipticals \citep{2001ApJ...555L..91N} 
does not influence the conclusions based on global kinematics of early-type merger remnants 
presented here.  
We performed second generation binary mergers with mass ratios of $\eta =1$ 
and $\eta = 3$ between anisotropic (11AA, 31AA) and rotating
(11RR, 31RR) progenitors.
The low-mass progenitor in the unequal-mass mergers contained a 
fraction of $1/ \eta $ of the mass and 
particle number of the high-mass companion in each subcomponent and its size was scaled as
$\sqrt{1/\eta}$. Particle numbers were the same as in NB03, 
resulting in 360000 luminous and 640000 dark particles for 11AA. 

Test simulations indicated that the initial orientation of the
early-type progenitors is less important than for spiral-spiral mergers.
Therefore we decided to focus on a variation of the pericenter distances of the
mergers as an analysis of cold dark matter simulations by
\citet{2003astro.ph..9611K} has shown that impact parameters
of major mergers are larger than typically assumed in isolated simulations.
For every initial setup the galaxies approached each other on nearly
parabolic orbits with three pericenter distances of
$r_{\mathrm{sep}} = 0.5, 2, 6$ where the disk scale length of the more massive first generation 
spiral was $h_{d} =1$ (see NB03 for all details) resulting in 12 
remnants in total. The initial separation was
fixed to 30 length units. The simulations were performed
with the treecode VINE in combination with the special purpose
hardware GRAPE-5 \citep{2000PASJ...52..659K}. The gravitational Plummer-softening 
length was the same ($\epsilon_{\mathrm{plum}} = 0.05$) as for the first generation 
spiral mergers. The particle orbits were integrated with a leapfrog integrator
and a fixed time-step of $\Delta t = 0.02$ which corresponds to
$\approx$ 0.2\% of the half-mass rotation period of the first
generation disks.  The total energy was conserved better than 
0.6\%. Every remnant was analyzed 10 dynamical timescales after the
merger was complete.   

The isophotal shape and the kinematics of 500 random
projections of each of the 12 merger remnants were analyzed. 
For each projection the characteristic ellipticity $\epsilon_{\mathrm{eff}}$ 
and isophotal shape parameter $a4_{\mathrm{eff}}$ as well as the rotational
velocity along the major axis, $v_{\mathrm{maj}}$, and the central velocity 
dispersion, $\sigma_0$, was determined as in NB03. 

The left panel of Fig.\ref{fig1} shows the
location of the remnants in the $({v_{\mathrm{maj}}/\sigma_0})^*-a_4$ plane.  
For every merger pair the projected remnants of the three  orbital 
geometries have been combined.
All 11AA (solid line), 11RR (dashed line) and 31AA (dotted line) remnants 
have $({v_{\mathrm{maj}}/\sigma_0})^* < 0.5$ and show 
predominantly boxy isophotes. Only 31AA remnants with a large pericenter 
distance, $r_{\mathrm{sep}} = 6$ and 11RR remnants with a small pericenter distance 
$r_{\mathrm{sep}} = 0.5$ can appear significantly disky ($a4_{\mathrm{eff}} \ge 0.5$). 
31RR remnants show distinct properties and have $({v_{\mathrm{maj}}/\sigma_0})^* \ge 0.5$  
with predominantly disky isophotes.  

The distinct kinematical properties of the merger remnants becomes 
more evident in the $({v_{\mathrm{maj}}/\sigma_0})-\epsilon_{\mathrm{eff}}$-plane
which is shown in the right panel of Fig.\ref{fig1}.
11AA, 11RR, and 31AA remnants are slow rotators in good agreement with observed 
boxy and anisotropic ellipticals (filled squares).
31RR remnants show stronger rotation 
($0.2 < {v_{\mathrm{maj}}/\sigma_0} < 0.5$) and are located in the 
overlap region between boxy and disky ellipticals. This analysis indicates that 
unequal mass mergers of early-type galaxies can result in slowly rotating anisotropic 
systems if the more massive progenitor had similar properties.  

\begin{figure}
\epsscale{0.8}
\plotone{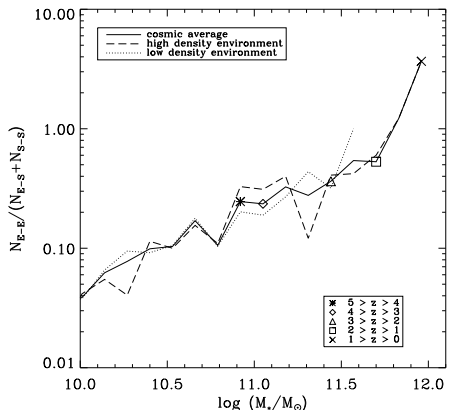}
\figcaption{Number ratio of last major mergers between two elliptical galaxies 
(E-E) and the sum of mergers between two spirals (S-S) and a spiral and an 
elliptical (E-S) as a function of stellar mass. The solid line shows the global
average while the dotted  and long dashed lines show results field- and 
high density (cluster) environments, respectively. The symbols show the maximum 
possible remnant mass formed by E-E mergers at different redshifts for 
which S-S and E-S merger still occur. 
\label{fig2}}
\end{figure}

\begin{figure}
\epsscale{0.8}
\plotone{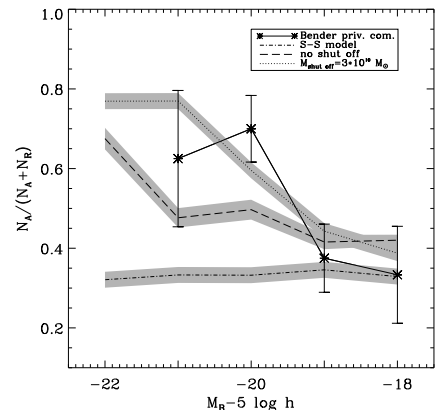}
\figcaption{Ratio of anisotropic ellipticals to the total number of elliptical galaxies 
as a function of their $B-$band luminosity $M_B$. Stars show observational data provided by 
Ralf Bender. The dot-dashed line shows the result for S-S mergers alone which can not reproduce 
the observed trend. The SAM predictions including early-type mergers without 
(dashed line) and with (dotted line) assuming a shut-off mass scale follow the 
observed trend. The shaded areas indicate the scatter in the SAMs. 
\label{fig3}}
\end{figure}

\section{Significance of early-type mergers and the mass dependence of elliptical
galaxy properties} 
\label{SEMI}
Semi-analytical models are useful to quantify the importance of early-type 
mergers for elliptical galaxy formation and to compare the theoretical models
with observations. The model used here is described in detail in KB05. 
For the present study we adopt the following cosmological parameters: $\Omega_0=0.3$, 
$\Omega_{\Lambda}=0.7$, $h=0.65$ and $\sigma_8=0.9$. 
Early type galaxies were defined by a bulge to total mass fraction of 
$ \geq 60 \%$. All galaxies with smaller bulge fractions were defined 
as spirals.  As the last major merger determines the final properties of the 
elliptical galaxies we show in Fig.\ref{fig2} the predicted number 
ratio of last elliptical-elliptical (E-E) mergers to the sum of the last 
elliptical-spiral (E-S) and the last spiral-spiral (S-S) mergers found in our 
simulations. The importance of E-E mergers steadily increases with 
increasing remnant mass. Remnants with stellar masses 
$M_* \gtrsim 6 \times  10^{11} M_{\odot}$ 
are generally assembled by E-E mergers. Our simulations predict
that the $N_{E-E}/(N_{E-S}+N_{S-S})-M_*-$relation  shown 
in Fig.\ref{fig2} is independent of redshift and environment. 
The only difference is that ellipticals reach larger masses
in denser environments and that at higher redshifts the relation 
stops at lower remnant masses. 

A summary of the conditions, which are based on the numerical simulations presented here and 
in NB03, 
that lead to anisotropic systems or systems with additional rotation is shown in Tab. 
\ref{tab1} where S denotes spiral galaxies. For example the combination 1R-3A denotes
a merger of a rotationally flattened elliptical with an anisotropic elliptical that is 
3 times more massive. This merger is assumed to lead
to an anisotropic (A) early-type galaxy. All simulations so far show that equal-mass 
spiral-spiral or elliptical-elliptical
mergers predominantly lead to slowly rotating anisotropic systems. It is therefore 
very likely that this will also be true for equal-mass mixed mergers between 
spirals and ellipticals. In the case of unequal-mass mixed mergers we make the 
conservative assumption that the property of the more massive
progenitor dominates the structure of the remnant. We now can classify all the
merger remnants in our semi-analytical models using the approach of KB05 and Table \ref{tab1}. 
The result (dashed line) is compared to observations kindly provided 
by Ralf Bender \footnote{extended data set based on 
\citet{1992ApJ...399..462B}} in Fig.\ref{fig3}. As anisotropic
 systems also form in some unequal-mass early-type mergers
the observed trend with galaxy luminosity can be reproduced.
An even better agreement (dotted line) is found if gas infall and star formation 
is suppressed  above a critical bulge mass of
$M_* \geq 3 \times 10^{10} M_{\odot}$, motivated by recent
predictions of e.g.  \citet{2005ApJ...625..621B} and \citet{2005ApJ...625...23B}. 
In this case, the growth of stellar disks by gas infall and star formation 
is suppressed above this critical mass limit, increasing the frequency
of early-type mergers and by this enhancing the formation rate of anisotropic 
systems. A model which generalizes the S-S results of NB03
to all mergers (dot-dashed line) fails to reproduce the observed trend.


\section{Conclusions}
\label{CONCLUSIONS}
Binary major mergers of spiral galaxies alone do not provide a viable formation 
mechanism for the whole population of giant elliptical galaxies. In particular 
they fail to reproduce the dominance of anisotropic, slowly rotating 
and boxy ellipticals at high luminosities. A revised major merger scenario 
which in addition allows major mergers of gas poor early-type galaxies 
helps to solve this problem. Individual numerical simulations of 
collisionless binary early-type mergers presented in this letter reveal 
an additional channel for the formation of anisotropic, 
slowly rotating ellipticals. In the context of hierarchical cold dark 
matter cosmologies this additional formation mechanism can successfully reproduce 
the observed trend that more luminous giant ellipticals are slower rotators and 
more supported by anisotropic velocity dispersions alone. A trend that matches 
observations nicely, if residual gas infall, star formation (e.g. the 
subsequent formation of disks) for galaxies with stellar bulge 
masses $M_* \geq 3 \times 10^{10} M_{\odot}$ is suppressed. A threshold value like 
this has been proposed recently \citep{2005ApJ...625..621B,2005ApJ...625...23B}. 
We can also conclude that dissipation becomes less important for the assembly of 
more massive ellipticals as proposed by \citet{1996ApJ...464L.119K}.  

The simulations also indicate that early-type merger remnants do preferentially 
have boxy isophotes (see also \citealp{1993MNRAS.261..379G}). However, individual merger 
geometries also lead to disky and anisotropic remnants. Further 
investigations will show whether this higher order 
effect requires the inclusion of additional physics like e.g. the coalescence of 
massive black holes. 

Our analysis strongly supports a scenario where major mergers between ellipticals 
play an important role for the assembly of elliptical galaxies even at low redshifts. 
Provided that the stellar population of the progenitor galaxies was old 
even the late assembly of an elliptical 
galaxy in a major early-type merger will not change this property. Further investigations 
will have to show in how far this 
scenario is in details consistent with the redshift evolution of ellipticals and local 
scaling relations like the color-magnitude relation, the Fundamental Plane and the 
black-hole-mass-$\sigma$ relation.  

\acknowledgments
We thank the referee for valuable comments on the manuscript. We are grateful to
 Eric Bell, Roland Jesseit and Fabian Heitsch for helpful discussions. This work 
was supported by the DFG Schwerpunktprogramm 1177. SK acknowledges 
funding by the PPARC Theoretical Cosmology Rolling Grant.

\end{document}